\RequirePackage{fix-cm}
\documentclass[smallextended]{svjour3}       
\smartqed  
\usepackage{graphicx}
\usepackage{booktabs}
\usepackage{natbib}
 \bibpunct[, ]{(}{)}{,}{a}{}{,}%
\setcitestyle{authoryear,round}
\usepackage[tight,footnotesize]{subfigure}
\usepackage{url}
\usepackage{doi}
\usepackage{algorithm}
\usepackage{algorithmic}

\usepackage{color}
\usepackage{stfloats}
\usepackage{amssymb}
\usepackage{amsmath}
\begin{document}\sloppy

\title{FedDPGAN: Federated Differentially Private Generative Adversarial Networks Framework for the Detection of COVID-19 Pneumonia
}


\author{Longling Zhang\textsuperscript{$\dagger$} \and Bochen Shen\textsuperscript{$\dagger$} \and Ahmed Barnawi \and Shan Xi \and Neeraj Kumar\textsuperscript{**} \and Yi Wu\textsuperscript{*}
         \thanks{\textsuperscript{$\dagger$}Equal contributions}
         \thanks{\textsuperscript{*}First corresponding author}
         \thanks{\textsuperscript{**}Second corresponding author}
}


\institute{Longling Zhang, Boshen Shen, Shan Xi, and Yi Wu \at
              School of Data Science and Technology, Heilongjiang University, Harbin, China, 150080 \\
              \email{20185759@s.hlju.edu.cn,20194983@s.hlju.edu.cn,2191831@s.hlju.edu.cn,1995050@hlju.edu.cn}           
           \and
           Ahmed Barnawi \at
              King Abdul Aziz University, Riyadh 11543, Saudi Arabia\\
              \email{ambarnawi@kau.edu.sa}
              \and
              Neeraj Kumar \at
              Thapar Institute of Engineering and Technology Pariala India\\
              \email{neeraj.kumar@thapar.edu}
}

\date{Received: date / Accepted: date}

\maketitle

\begin{abstract}
Existing deep learning technologies generally learn the features of chest X-ray data generated by Generative Adversarial Networks (GAN) to diagnose COVID-19 pneumonia. However, the above methods have a critical challenge: data privacy. GAN will leak the semantic information of the training data which can be used to reconstruct the training samples by attackers, thereby this method will leak the privacy of the patient.  \textcolor{black}{Furthermore, for this 
reason that is the limitation of the training data sample, different hospitals jointly train the model through data sharing, which will also cause the privacy leakage. }
\textcolor{black}{To solve this problem, we adopt the Federated Learning (FL) framework which is a new technique being used to protect the data privacy. Under the FL framework and Differentially Private thinking, we propose a Federated Differentially Private Generative Adversarial Network (FedDPGAN) to detect COVID-19 pneumonia for sustainable smart cities.}
 Specifically, we use DPGAN to privately generate diverse patient data in \textcolor{black}{which differential privacy technology is introduced to make sure the privacy protection of the semantic information of training dataset.} Furthermore, we leverage FL to allow hospitals to collaboratively train COVID-19 models without sharing the original data. 
 \textcolor{black}{Under Independent and Identically Distributed (IID) and non-IID settings, The evaluation of the proposed model is on three types of chest X-ray (CXR) images dataset (COVID-19, normal, and normal pneumonia). A large number of the truthful reports make the verification of our model can effectively diagnose COVID-19 without compromising privacy.}
\end{abstract}
\section{Introduction}
\textcolor{black}{COVID-19 is a highly contagious infectious disease which is caused by Severe Acute Respiratory Syndrome Coronavirus-2 (SARS-CoV-2) in which it and its variants are still spreading worldwide \citep{cao2020covid}.} As COVID-19 is raging around the world, hospitals lack sufficient staff to diagnose suspected patients with COVID-19 disease. \textcolor{black}{To improve the efficiency of diagnosing COVID-19, researchers generally develop deep learning systems to diagnose patients' chest X-ray images \citep{liang2020early,ting2020digital,chang2018computational}.} \textcolor{black}{\cite{dhiman2021adopt} proposed the J48 decision three model in order to discover the COVID-19 samples and gained the conclusion that the method is useful.} \textcolor{black}{Although deep learning technologies have greatly improved the efficiency of diagnosis, these technologies base on a great amount of annotated instances and labeled data, which is hard for hospitals to find enough training samples \citep{wang2020covid,abdel2021fss,abdel2020hsma_woa,abdel2021intelligent}.} \textcolor{black}{Furthermore, due to the privacy of medical image dataset, hospitals find it hard to gather enough samples.} Therefore, the above data availability issues have brought many severe challenges to such a diagnostic method.

Inspired by the Generative Adversarial Networks (GANs) techniques \citep{goodfellow2014generative}, researchers generally use GAN to generate diverse training data to solve the data availability issues. \textcolor{black}{For example, \cite{waheed2020covidgan} proposed a GAN model to generate CXR images by developing an ACGAN model.} \cite{bao2020covid} proposed a model named COVID-GAN to predict the impact of COVID-19 epidemic. However, the above-mentioned method using GAN as a generative model will leak the patient's private information. \textcolor{black}{Since the GAN-based model remembers the semantic information of the distribution of a big deal number of training samples, it is easy for a hacker to use reverse engineering to reconstruct the patient's private data \citep{jordon2018pate,liu2019ppgan,xu2019ganobfuscator,chen2020gan}.} For instance, \cite{gu2020image} used reverse engineering technique to obtain the hidden encoding of the real image combining the corresponding feature maps of the middle layer generated in the generator and then reconstructed picture similar to the real image. \textit{Therefore, we need to seek a way to design a data generative model that does not reveal privacy.}

Furthermore, researchers also utilize data sharing strategies to solve data availability issues \citep{cosgriff2020data}. \cite{moorthy2020data} designed a data sharing strategy to enable the hospital to have enough labeled data to train deep learning models. Data sharing methods have been used by hospitals to expand the scale of training samples until the issuance of the General Data Protection Regulation (GDPR) \citep{voigt2017eu}. The reason why we can no longer use data sharing methods is that the GDPR stipulates that organizations do not allow arbitrary sharing of user data, because this will leak user privacy. In particular, COVID-19 data is medical data, which is very sensitive to patients. \textit{Therefore, we need to seek novel learning methods to avoid data sharing that would leak privacy and violate the law.}

\textcolor{black}{First, since Differential Privacy (DP) technology is generally used in privacy protection,} previous works \citep{xie2018differentially,liu2019ppgan,xu2019ganobfuscator,jordon2018pate,hitaj2017deep} focused on using DP technology to alleviate the problem of GAN leaking privacy. For example, \cite{xie2018differentially} and \cite{liu2019ppgan} proposed Differentially Private GAN (DPGAN) to protect the user privacy by leveraging $(\varepsilon,\delta)$-DP technique. However, such DPGAN-based 
models are only suitable for centralized learning rather than distributed learning. Second, due to GDPR's restrictions on data sharing strategies, data exists between hospitals in the form of ``islands" \citep{9205981,li2020federated}, which inspired researchers to develop a privacy-persevering distributed machine learning paradigm, i.e., Federated Learning \citep{mcmahan2017communication}. In this context, references \citep{li2019privacy,ge2020fedner,sheller2020federated,sui2020feded} applied FL in medical fields to develop some privacy-persevering applications such as Medical Imaging \citep{li2019privacy}, Medical Relation Extraction \citep{sui2020feded}, and Medical Named Entity Recognition \citep{ge2020fedner}.

We gain inspiration in these above methods and propose the Federated Differentially Private Generative Adversarial Network (FedDPGAN) model to detect COVID-19 pneumonia without compromising patient privacy. In this model, DPGAN is a key component of the proposed model and its function is by adding Gaussian noise in training gradient that protects the training samples' privacy. In particular, we introduce a federated learning framework and developed a distributed DPGAN to enable different hospitals to train COVID-19 diagnostic models collaboratively without data sharing. Specifically, with the help of FL and its aggregation mechanism, FedDPGAN can aggregate model parameters from medical institutions in different geographical locations to construct a global GAN model with well-preserved privacy. \textcolor{black}{A large number of truthful data studies that FedDPGAN model is better than the existing centralized learning and FL-based models. The research contents and contributions include as:}

\begin{itemize}
	\item \textcolor{black}{Unlike existing frameworks, we propose a Federated Differentially Private Generative Adversarial Network framework, which enables different hospitals can utilize the privacy-preserving data augmentation method, i.e., distributed DPGAN model to generated high-quality training samples which relieves the problem of lacking the training sample of COVID-19 then apply ResNet \citep{he2016deep} model in FL to achieve high-precision COVID-19 detection.}
	
	\item \textcolor{black}{To address the data availability issues in detecting COVID-19, we design a distributed DPGAN by leveraging FL framework. In particular, we find that distributed DPGAN can alleviate the Non-independent and identically distributed (non-IID) issue in FL. Specifically, with the help of FL and its aggregation mechanism, FedDPGAN can construct a global and local data augmentation model by aggregating model parameters from medical institutions in different geographical locations to do different medical task.}
	
	\item \textcolor{black}{We conduct extensive case studies on different pneumonia CXR images demonstrate that the proposed model FedDPGAN is better than the existing centralized learning and FL-based models. Specifically, our model shows that the best centralized model by 1.52\% and the FL model by 0.49\% in IID distribution. In non-IID data distribution, our model performance is 3.00\% higher than the best FL model.}
	
\end{itemize}

\textcolor{black}{The article organizational structure is as follows. Section II is literature review about differential Private Generative Adversarial Networks and Federated Learning. Section III Section IV presents the FedDPGAN algorithm. In FedDPGAN, we applicate the FedAVG algorithm and differential private method makes introduction particularly. The final result compares between Sections V and V-F. The Section VI includes brief summary of the article.}

\section{Related Work}
In this section, we summarize the advanced work in the Differentially Private Generative Adversarial Networks (DPGAN) and Federated Learning (FL) fields.

\subsection{Differentially Private Generative Adversarial Networks}

How to combine deep learning technology with privacy protection technology is an emerging research direction. \textcolor{black}{For example, many researchers apply differential privacy (DP) technologies to training model that ensures models security. \cite{abadi2016deep} developed a privacy-preserving deep learning model training paradigm by adding well-designed differential privacy noise (i.e., Gaussian noise) when computing the model gradients.} \cite{voigt2017eu} proposed a DP-based deep learning models to achieve privacy-preserving disease classification application. In particular, \textcolor{black}{\cite{xie2018differentially} suggested using DPGAN to protect the user privacy by leveraging $(\varepsilon, \delta)-\mathrm{DP}$ techniques.} Inspired by the above work, the current work focuses on using DPGAN to develop some medical-related applications \citep{choi2017generating,xie2018differentially,Chang_2020_CVPR}. \cite{Chang_2020_CVPR} utilized DPGAN to develop a medical imaging application.

However, such DPGAN-based models are only suitable for centralized learning rather than distributed learning.\textcolor{black}{As a result, although researchers have solved the above problems and proposed some privacy preserving methods to protect the local model training, distributed DPGAN has not been studied yet. We propose the distributed DPGAN that can be applied in distributed learning-based applications.}

\subsection{Federated Learning}

\textcolor{black}{Federated Learning (FL) \textcolor{black}{\citep{mcmahan2017communication} will establish a data protection model,} distributing dataset on each client machine, and aggregating locally-computed updates for a globally model which helps the participating clients to achieve experimental results similar to distributed data \citep{9170265,liu2020towards}, while maintaining the privacy of the training data \citep{9184079}.} Therefore, as a promising distributed machine learning framework for privacy protection, FL has spawned many emerging applications such as Google Keyboard \citep{hard2018federated}, traffic flow prediction \citep{9082655}, anomaly detection \citep{9146846,wu2019dominant}, medical imaging \citep{sheller2020federated}, etc. In particular, medical institutions turn their attention to FL to develop a collaborative learning paradigm for privacy protection, thereby avoiding legal problems caused by data sharing. \textcolor{black}{For example, \cite{ge2020fedner} applied Medical Named Entity Recognition (NER) in FL by utilizing different hospitals data to promote the hospitals training models.} \cite{chen2020fl} proposed FL-QSAR model contributing the performance in QSAR prediction by collaborating among pharmaceutical institutions in drug discovery. However, the non-IID problem in FL hinders the rapid development of FL in the medical field.

To address this problem, many novel optimization algorithms are designed to overcome the adverse effects of non-IID. \textcolor{black}{\cite{9155494} designed a mechanism based on deep Q learning to maximize the reward for overcoming this problem by adopting the method of selecting a subset of devices during each round of communication.} However, such optimization algorithms are suitable for mobile IoT and cannot be applied in the medical field. The reason is that this algorithm requires complex client selection and training complex deep Q-leaning models. In this paper, we find that GAN can address this problem by generating diverse training samples for FL training. In short, solving the non-IID problem in FL is the only way to apply FL in the medical field.

\section{Preliminary}
\subsection{Differential Privacy}

DP is widely used to maintain secure model and protects the training data, hence it is a privacy protection technology. The classic definition of DP provided below relies on the concept of so-called adjacent databases, that is, databases that differ in only one element (or sample, as it is the case in Machine Learning datasets). Therefore, the formal definition of DP is as follows:

\begin{definition}[$(\varepsilon, \delta)-DP$ \citep{dwork2014algorithmic}] \label{defi-1} \textcolor{black}{Any two adjacent datasets $b$ and $b^{\prime}$ are input, with the random algorithm $\mathcal{K}$ and the subset of outputs $S$ hold that:}
	\begin{equation} 
	\textcolor{black}{
		\Pr [{\mathcal{K}}(b) \in S] \le {e^\varepsilon } \cdot \Pr \left[ {{\cal M}\left( {{b^\prime }} \right) \in S} \right] + \delta,
		}
	\end{equation}
	\textcolor{black}{where $\epsilon$ is the privacy budget and $\delta$ is the failure probability. The smaller the $\delta$, the closer the distribution of the data output by $\mathcal{K}$ in $b, b’$ datasets.}
	
\end{definition}
\textcolor{black}{
According to Definition \ref{defi-1}, we can use DP to ensure the privacy of the semantic information of training data.} But DP can not be directly applied in deep learning because DP is applicable to query functions in the database. Therefore, researchers generally apply DP in deep learning by adding the well-designed Gaussian noise that is in keeping with the definition of the differential privacy. The formal definition of Gaussian noise mechanism is as follows:

\begin{definition} [Gaussian Noise Mechanism] \label{defi-2} \textcolor{black}{$S_f$ is a random function 
sensitivity for the two adjacent dataset $b, b’$, $f(b)$ is the query function, and ${S_f^2 \cdot {\sigma ^2}}$ is the variance of the Gaussian distribution. For each pair of adjacent inputs $b$ and $b'$, the Gaussian noise mechanism can be expressed as follows:}
	\begin{equation}
	\textcolor{black}{
		\mathcal{M}(b) \triangleq f(b) + \mathcal{N}\left( {0,S_f^2 \cdot {\sigma ^2}} \right),
		}
	\end{equation}
	\textcolor{black}{
	where $\mathcal{N}\left(0, S_{f}^{2} \cdot \sigma^{2}\right)$ is the noise to disturb the distribution in 0 and standard deviation ${S_f} \cdot \sigma $. Then we give the definition of the sensitivity $S_f$ of the random function $f$ as follows:}
\end{definition}

\begin{definition} [Sensitivity \citep{mironov2017renyi}] \label{defi-3}The sensitivity of the random function $f$ is as follows:
	\begin{equation}
	\textcolor{black}{
		\Delta f = \mathop {\max }\limits_{b,b'} {\left\| {f(b) - f\left( {{b^\prime }} \right)} \right\|_2},
		}
	\end{equation}
	\textcolor{black}{where prioritize the two adjacent datasets $b, b^{\prime}$.}
\end{definition}

According to Definition \ref{defi-2} and Definition \ref{defi-3}, it can be seen that the core meaning of the parameter of sensitivity indicating the magnitude of noise is to indicate the effect of deleting deleted records in the data set on the query result. That is to say, the noise scale change of the Gaussian noise mechanism is proportional to the sensitivity. In particular, when $\delta  \geqslant \frac{4}{5}\exp \left( { - {{(\sigma  \cdot \varepsilon )}^2}/2} \right)$ and $\varepsilon<1$, the random function $f$ satisfies the definition of $(\varepsilon, \delta)$-DP after adding Gaussian noise.

\subsection{Generative Adversarial Networks}

\textcolor{black}{Generative Adversarial Network (GAN) is an approach of the non-supervision model. GAN contains two parts: Generator $N$ and Discriminator $M$. The generator randomly takes samples which is from potential space (latent space) and emulate the truthful traning data more and more. Input set of discriminator network is truthful output data, that  distinguish training data from truthful samples as much as possible.}  \textcolor{black}{Inspired by Game Theory, the generative model $N$ make effort that capture distribution of the data, and the  model to discrimate $M$ tries to estimate the probability.} They confront and constantly adjust the parameters during training. The ultimate goal of GAN is to make the discriminator unable to judge whether the output result of the generator is true or fake. We mentioned above that the GAN optimization problem is actually a game theory of $N$ and $M$, that is, a minimal-maximization problem, so reference \citep{goodfellow2014generative} proposed an important approach is to solve this problem:

\begin{definition} [Optimal Generator] \textcolor{black}{For generator, it learn a distribution ${P_{g}}$ of dataset. The input data distribution ${P_{z}(z)}$, the generator $G\left(z ; \theta_{g}\right): z \rightarrow x $, and the discriminator $G\left(z ; \theta_{g}\right): z \rightarrow x$.} Therefore, the optimal generator of a function can be expressed as follows:
	\begin{equation}
	\textcolor{black}{
		\begin{split}
			\min _{N} \max _{M} V(M, N) =&E_{x \sim P_{\text {data }}(x)}[\log M(x)]+\\
			&E_{z \sim P_{z}(z)}[\log (1-M(N(z)))].
		\end{split}
		}
	\end{equation}
\end{definition}

However, the above optimization model has the problems of vanishing gradient and samples diversity. Therefore, researchers put forward an optimized GAN, which solves the problem of gradient disappearance, which is defined as follows:
\begin{definition}[Optimize GAN] \textcolor{black}{$\prod\left(\mathrm{F}_{\gamma}, \mathrm{F}_{g}\right)$ is the set of joint distribution $\gamma$ include $\mathrm{F}_{\gamma}$ and $\mathrm{F}_{g}$ all possible combinations. $\mathrm{F}_{\gamma}$ and $\mathrm{F}_{g}$ are the edge distribution.}
	
	\begin{equation}
	\textcolor{black}{
		W\left(\mathrm{~F}_{\gamma}, \mathrm{F}_{g}\right)=\inf _{\gamma \in \prod\left(\mathrm{F}_{\gamma}, \mathrm{F}_{g}\right)} E_{(a, b) \sim \gamma}[\|a-b\|]
		}
	\end{equation}
\end{definition}

\subsection{Federated Learning}
As a promising distributed machine learning framework for privacy protection, Federated Learning protects users' privacy data by keeping their local data locally and only periodically exchanges updates with the server which reduces their communication costs. The classic algorithm for optimizing federated optimization problems is Federated Averaging (FedAvg) \citep{mcmahan2017communication}.

In FL, we consider a server $\mathcal{S}$ and a subset of the clients $\mathcal{K}$ participating in the training of a shared global model $F(\cdot)$. We assume that each client holds an IID or non-IID datasets ${D_k}$. At the client side for data sample $x$, we let $\ell(\omega ; x)$ be the loss function, where $\omega  \in {\mathbb{R}^d}$ denotes the model's trainable parameters. At the server side, we let $\mathcal{L}(\omega)=E_{x \sim \mathcal{D}}[\ell(\omega ; x)]$ be the loss function and let the server to optimize the following objective function:
\begin{equation}
	{\min _\omega }\mathcal{L}(\omega ),{\text{ where }}\mathcal{L}(\omega ): = \sum\limits_{k = 1}^K {{p_k}} {\mathcal{L}_k}(\omega ),
\end{equation}
\textcolor{black}{where $K$ represents the clients}, $({p_k},\sum\limits_k {{p_k}}  = 1)$ indicates the relative influence of each client on the global model. In FL, the training is conducted between the server and clients side in a $T$-round communication rounds to minimize the above objective function following a three step protocol:
\begin{itemize}
	\item  \textbf{\textit{(Step 1, Initialization):}}  \textcolor{black}{The $t$-th round of training, the server selects a subset from clients $\mathcal{K}$ to participate in training and broadcasts the initialized global model parameters $\omega^{t}$ to each client.}
	
	\item \textbf{\textit{(Step 2, Local Training):}}  Each client individually executes the local training to obtain the model updates. \textcolor{black}{ Specifically, the client trains the received global model $\omega^{t}$ on the dataset $\mathcal{D}_{k}$ by using the local optimizer, e.g., Stochastic Gradient Descent (SGD) and then uploads all updates $\Delta \omega_{k}^{t}$ to the server.}
	
	\item \textbf{\textit{(Step 3, Aggregation):}} After collecting all updates uploaded by $\mathcal{K}_{t}$ clients, the central server uses the aggregation algorithm, i.e., \textcolor{black}{FedAvg Algorithm to obtain the new global model which serves as an initial point for the next communication round by aggregate the model updates.}
\end{itemize}

\textcolor{black}{Repeat the above steps until the global model converges.}


\begin{figure*}
	
	\includegraphics[width=12cm]{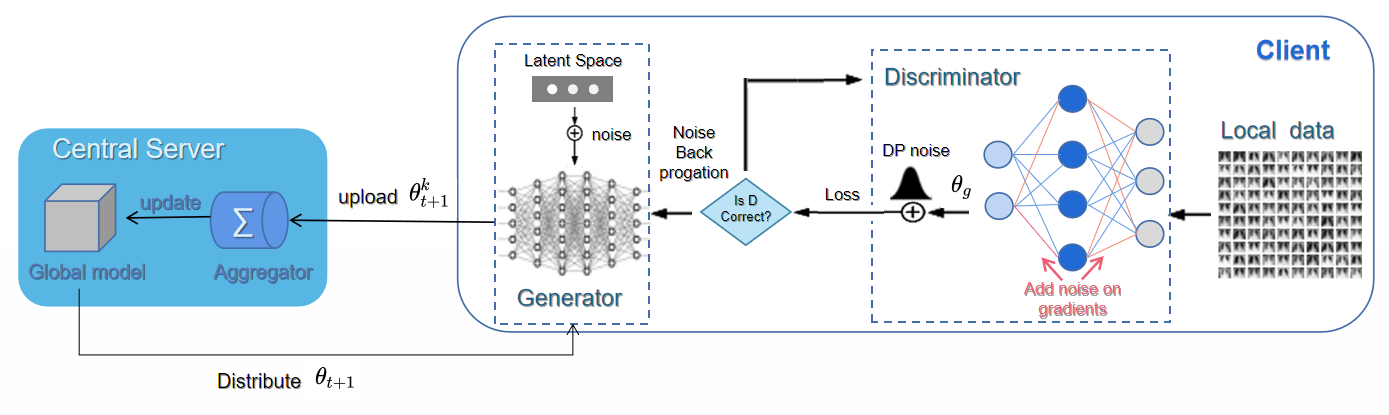}
	\centering
	\caption{Overview of Federated  Differentially Private Generative Adversarial Network (FedDPGAN) framework.}\label{fig:picture}
\end{figure*}

\section{PROPOSED FedDPGAN MODEL}
\textcolor{black}{We start with an introduction of medical DPGAN model. Then we introduce the FedDPGAN to diagnosis COVID-19 CXR images from various medical platforms. Specifically, FedDPGAN is a client-server architecture in which server shares the global model and coordinates the client's local privacy protection DPGAN model with SGD optimizer.}

\subsection{Architecture of FedDPGAN}
In this subsection, we present the approaches to protect the users' private data, including the basic DP noise mechanism and FedDPGAN algorithm. In our framework, there is a central server $\mathcal{S}$ and a client set $\mathcal{K}$ with their local dataset $\mathcal{D}_k$. Next, we introduce in detail the functions of the components of the proposed model.

\subsubsection{Distributed DPGAN}
First, we present our distributed DPGAN mechanism. Since COVID-19 data is very private, we need to protect the privacy of patients when we use GAN to generate COVID-19 data. Therefore, following existing works \citep{liu2019ppgan,xie2018differentially}, we adopt the method of adding Gaussian noise to the training gradient to ensure dataset security.  Specifically, we adopt this way by adding random noise in discriminator which interferes with original data distribution, thereby protecting the privacy of the training data. Therefore, according to Definitions \ref{defi-1}--\ref{defi-3} we have:
\begin{equation}
	{g_\sigma } \leftarrow {g_\sigma } \cdot min(1,\frac{C}{{\left\| {{g_\sigma }} \right\|}}) + \mathcal{N}(0,\sigma _n^2c_g^2I),
\end{equation}
where $g_\sigma$ is the noisy gradient, $c_g$ is the sensitivity of the gradient function and $C$ is the clipping threshold. The gradient as a random variable approximately obeys Gaussian distribution, we have:
\begin{equation}
	f(x) = \frac{1}{{\sqrt {2\pi \sigma } }}\exp \left( { - \frac{{{{(x - \mu )}^2}}}{{2{\sigma ^2}}}} \right) \cdot \varepsilon ,
\end{equation}
where $\mu$ is the mean of the random variable.

To make the above random function $f$ satisfy the definition of differential privacy, we assume that $\Delta f = \mathop {\max }\limits_{d,d' \in \mathcal{D}} {\left\| {f(x) - f(x')} \right\|_2}$ is the $L2$-Sensitivity and $\forall \delta  \in (0,1)$, we have:
\begin{equation}
	\sigma  > \frac{{\sqrt {2\ln (1.25/\delta )} \Delta f}}{\varepsilon }.
\end{equation}
When we add Gaussain noise $ \mathcal{N}\left(0, \sigma_{n}^{2} c_{g}^{2} I\right)$ to the gradient, If and only if
\begin{equation}
	{\sigma _n} = 2q\sqrt {{n_d}\log \left( {\frac{1}{\delta }} \right)} /\varepsilon ,
\end{equation}
\textcolor{black}{where sample rate is $q$. we can say that this function $f$ satisfies the definition of DP.} To prevent the gradient from exploding, we use gradient clipping technology to make the gradient in the range of $(-c, c)$.

\subsubsection{FedDPGAN Framework}
Following the client-server architecture in FL, \textcolor{black}{the central server $\mathcal{S}$ selects a random fraction $C$ of clients. Then the clients aggregate a model broadcasting the model to each client. After global model is initialized,} the client uploads generator model's parameter $\theta_{t+1}^{k}$ to the aggregator. Then the aggregator accumulates $\sum_{k=1}^{k} \frac{N_k}{N} \theta_{t+1}^{k}$ getting the average value of $\theta_{t+1}$ and updates to the global model. 
\begin{equation}
	\theta_{t+1} \gets \sum_{k=1}^{k} \frac{N_k}{N} \theta_{t+1}^{k}.
\end{equation}

The server coordinates multiple clients updating and shares a new global model into the clients. The specific steps between clients and server can be summarized in three steps:
\begin{itemize}
	\item  \textbf{\textit{(Step 1, FL Initialization):}} \textcolor{black}{Firstly,  central server picks the subset of the clients. Then it broadcasts the initialized generator parameter $\theta_t$.}
	
	\item \textbf{\textit{(Step 2, Distributed DPGAN Training):}} After the initialization, the selected clients $K$ perform training iterations of SGD over their local data. We add Gaussian noise $\mathcal{N}(0,\sigma_n^2c_g^2I)$ appropriately in a bounded range $min(1,\frac{C}{\left\|g_\sigma \right\|})$ and then it will automatically clipping parameters to add noise when the next time update. The client updates the weight parameters $\omega$ and truncating in the range of $(-c, c)$ after updating the weight parameters to optimize the discriminator.
	
	\item \textbf{\textit{(Step 3, FL Aggregation):}} The clients upload their model parameters $\theta_{t+1}^{k}$ to the aggregator. The aggregator aggregates all model parameters getting the average value of $\sum_{k=1}^{k} \frac{N_{k}}{N} \theta_{t+1}^{k}$ and then updates to server to create a global model that is used as an initialization point for the next communication round.
	
\end{itemize}

\begin{algorithm}[!t]
	\caption{Federated  Differentially Private Generative Adversarial Network (FedDPGAN) Algorithm.}
	\begin{algorithmic}[1]
		\REQUIRE
		$P$: The platform set;
		$\alpha$: The learning rate;
		$c$: The clipping parameter;
		$m$: The mini-batch size;
		${N_d}$: Discriminator iteration;
		${N_g}$: Generator iteration;
		${\sigma_n}$: Noise scale;
		${P_g(Z)}$: Noise prior;
		${P_{data}(X)}$: Data generating distribution;
		\ENSURE
		$\theta$: DP generator;
		\STATE \textbf{Server executes:}
		\STATE Initialize generator parameters $\theta_0$ and discriminator parameters $\omega_0$;
		\FOR{each client $k \in P$ in parallel}
		\STATE $\theta_{t+1}^{k} \gets$ \textbf{ClientUpdate($\theta_t$, $\omega_t,k$)}\\
		\STATE $\theta_{t+1} \gets \sum_{k=1}^{k} \frac{N_k}{N} \theta_{t+1}^{k}$   
		\ENDFOR
		\STATE \textbf{ClientUpdate($\theta_t$, $\omega_t,k$):}\\
		\FOR{$t_1 = 1,...,N_g$}
		\FOR{$t_2 = 1,...,N_d$}	  
		\STATE {A mini-batch $\left\{Z_1,...,Z_m\right\}$ from $P_g(Z)$}\\
		\STATE {A mini-batch $\left\{X_1,...,X_m\right\}$ from $P_{data}(X)$}\\
		\STATE {$g_\omega \gets g_\omega$ min(1,$\frac{C}{\left\|g_\omega \right\|}) + N(0,\sigma_n^2c_g^2I)$\textbf{\textcolor{black}{(adding noise)}}}
		\STATE {$\omega \gets clip(\omega+\alpha\cdot SGD(\omega,g_\omega),-c,c)$}
		\ENDFOR
		\STATE {\textcolor{black}{$g_\sigma \gets g_\sigma$ min(1,$\frac{C}{\left\|g_\sigma \right\|}$)}}
		\STATE {$\theta \gets \theta - \alpha\cdot SGD(\theta,g_\theta)$}
			\ENDFOR
			\RETURN $\theta$ to server
		\end{algorithmic}
	\end{algorithm}
	
\subsection{FedDPGAN-based COVID-19 Diagnosis Model}
We represent the COVID-19 dataset characteristics and the COVID-19 diagnpsis model. First, we use the publicly available COVID-19 dataset as a benchmark dataset for evaluating the performance of the proposed model. This dataset consists of chest X-ray images of patients. To this end, we need to apply advanced Convolutional Neural Network (CNN) structure suitable for vision tasks in FedDPGAN to achieve higher performance.
	
Therefore, in this paper, we use ResNet \citep{he2016deep} model to diagnosis COVID-19 by classifying the chest X-ray images. Specifically, ResNet is a powerful emerging deep learning model that has attracted considerable attention in recent years. \textcolor{black}{ResNet adds a direct connection channel to the network structure to quickly transfer the training gradient, which greatly improves the efficiency of model training. Specifically, we get the gradient after the lower layer network training the parameters, the gradient is direct to transmit to the upper layer network parameters, that is, the original input information is allowed to be directly transmitted to the upper layer. Also, the correlation of gradients decays with the increase of layers. It has been proved that RESNET can effectively reduce the attenuation of this correlation. This feature enables ResNet to build a deeper network structure, which is widely used in image classification and is suitable for our COVID-19 medical image classification task.}
	
\section{EXPERIMENTAL RESULTS}
\textcolor{black}{We apply real-world CXR images that comprehensively evaluate the proposed model. First, we give details of the experimental environment, datasets, hyperparameters, and model details for this experiment. Second, we compare the other baseline model like
 advanced centralized learning model and electronic language-based model to determine the performance of our proposed model. Then, we compare the model performance with other benchmark models under simulated non-IID distributions. Finally, the influence of privacy parameters on model performance is discussed.}

\subsection{Experimental Setting}
\subsubsection{Datasets}
\begin{figure}
		\centering
		\subfigure[Real image]{
			\begin{minipage}[b]{.45\linewidth}
				\includegraphics[width=0.95\linewidth]{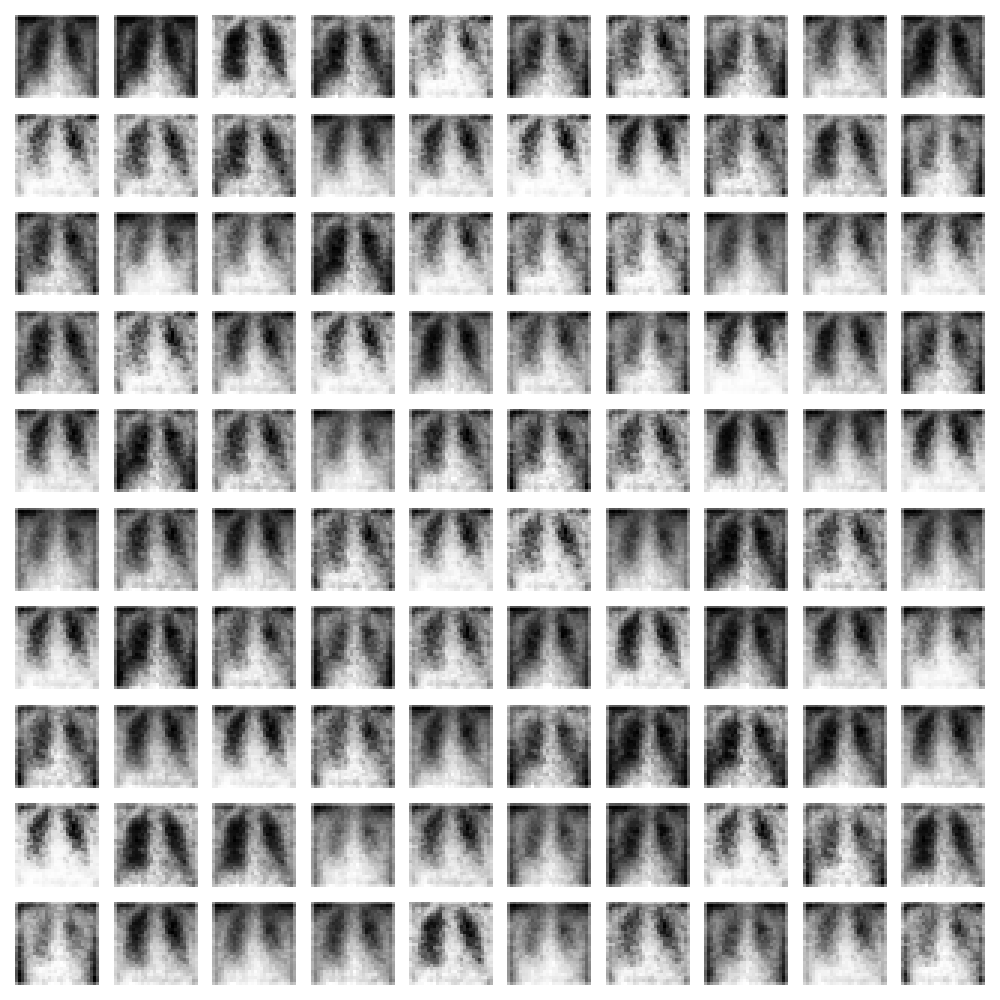}
			\end{minipage}
		}
		\subfigure[Fake image]{
			\begin{minipage}[b]{.45\linewidth}
				\includegraphics[width=0.95\linewidth]{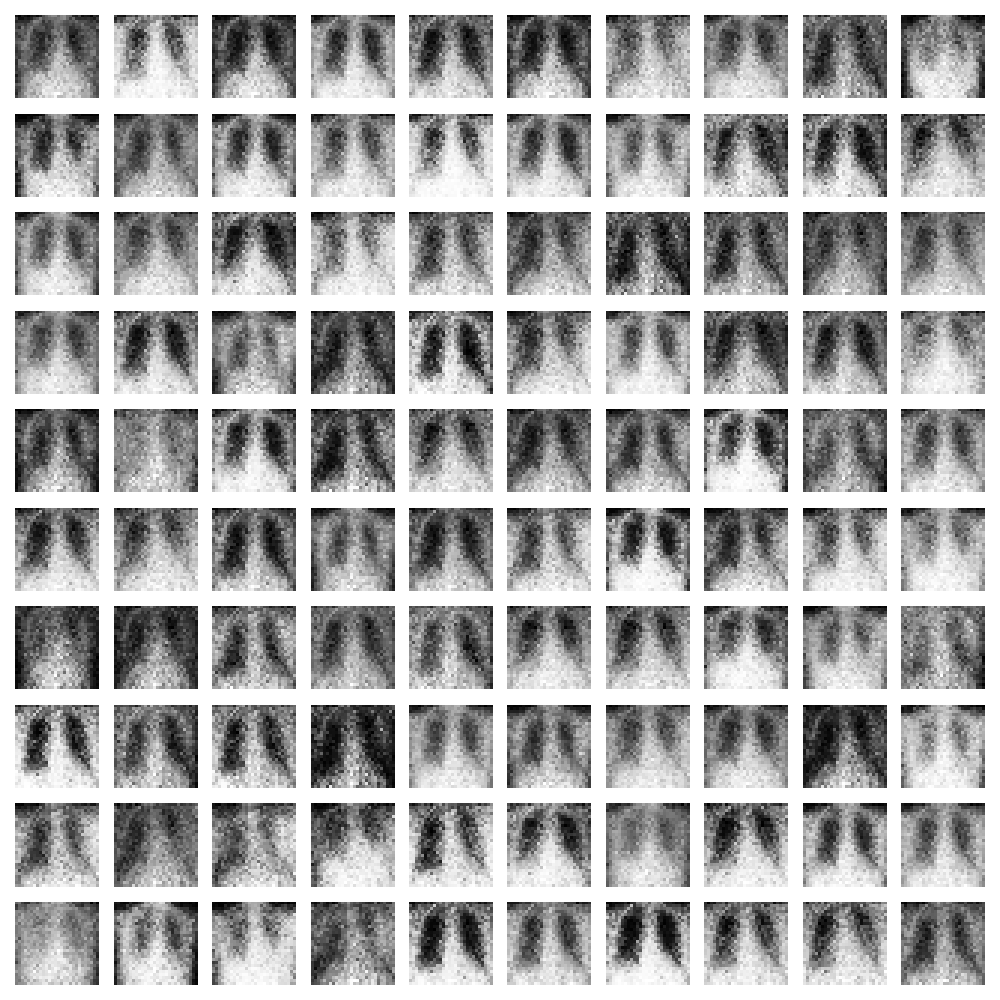}
			\end{minipage}
		}
		\caption{Overview of the generated dataset.}
		\label{fig:images}
	\end{figure}

\textcolor{black}{We evaluate the FedDPGAN on different pneumonia images dataset published by \cite{cohen2020covidProspective,cohen2020covid},} where the dataset consists of normal lung images, ordinary pneumonia images, and COVID-19 pneumonia images. Specifically, such a dataset contains 2,000 normal images, 1,250 normal pneumonia images and 350 COVID-19 pneumonia images. As mentioned above, we can find that this data has the problem of class imbalance, which is why we use the model DPGAN to generate diverse data. We generate fake chest X-ray images through DPGAN model and mix them into our dataset. More details can be seen in Fig.\ref{fig:images}. \textcolor{black}{In addition, we adjusted the image size about $28 \times 28$ pixels that speeds up the convergence of the model.}

\subsubsection{Baseline Models and Aggregation Method}
In this experiment, we use FedResNet, centralized ResNet \citep{he2016deep}, Convolutional Neural Network (CNN) \citep{tajbakhsh2016convolutional}, Multilayer Perceptron (MLP) \citep{li2018novel}, $K$ Nearest Neighbors (KNN) \citep{park2018privacy}, \textcolor{black}{and Support Vector Machines (SVM) \citep{morra2009comparison} as our baseline models that proves FedDPGAN performance.} Note that we apply ResNet model in our FedDPGAN framework.
	
Second, we use the FedAvg aggregation algorithm as the updated aggregation algorithm in the proposed framework. The reason is that the training model under the classic FedAvg aggregation algorithm performs well in various tasks.
	
\subsubsection{Non-IID Setting}
\begin{figure}
		\centering
		\large
		\includegraphics[width=0.8\linewidth]{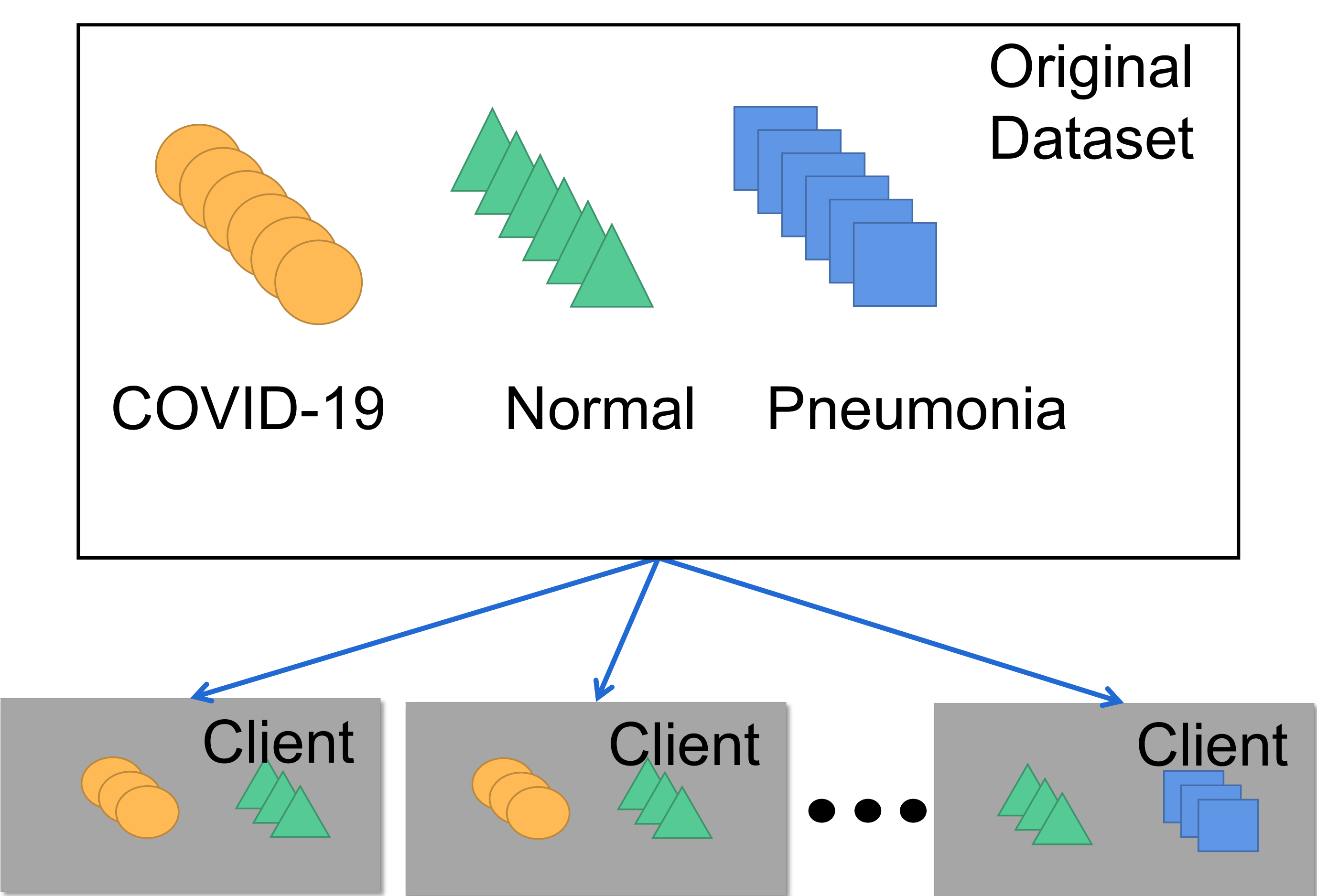}
		\caption{Overview of non-IID data allocation method.}
		\label{fig:NonIID}
	\end{figure}

In the medical field, since the data of different hospitals are collected by different types of collection equipment, the data between different hospitals is non-IID. In this paper, to achieve non-IID data distribution,  we assign two types of data, i.e., normal chest images and general pneumonia images to most clients and we put COVID-19 images into only a few clients. More details can be seen in Fig. \ref{fig:NonIID}.
	
\subsubsection{Hyperparameters}
In this experiment, we set $K=100$ clients and place the equal size dataset in each client. At each round of communication, we randomly select $C = 10\%$ of clients to participate in training and set the the local batch size $B = 10$, local epochs $E=5$, the learning rate of $\alpha = 0.01$, and  Gaussian noise generator generates the noise which default $\sigma = 0.0001$.
	
\subsubsection{Implementation and Setup}
\textcolor{black}{The implementation of the model is under the TensorFlow 2.0\citep{abadi2016tensorflow}, which is a powerful framework released by Google that can run on the GPU for acceleration. PyTorch\citep{paszke2019pytorch} is an open-source ML toolkit that hastens everything ranging from research prototyping to production deployment. All of the experiments are conducted using PyTorch and TensorFlow with Ubuntu 16.04. Experiments are conducted on a Linux Server with NVIDIA GeForce RTX 2080TI GPU and an i7 9900K CPU. }
	
	\begin{figure*}
		\centering
		\large
		\subfigure []{\includegraphics[width=0.45\linewidth]{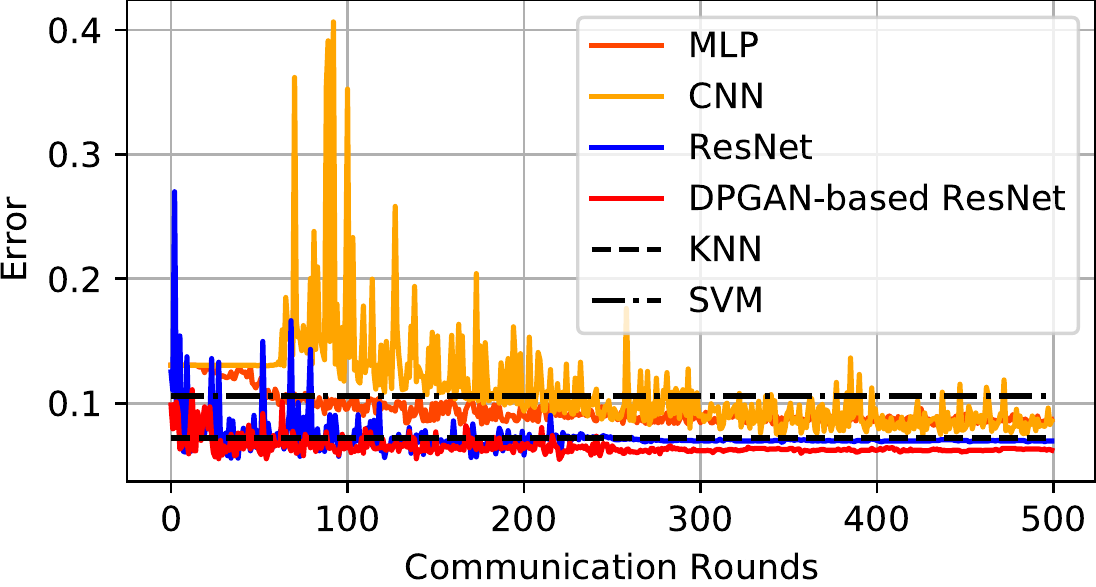}
			\label{fig3-a}}
		\hfill
		\subfigure[]{ \includegraphics[width=0.45\linewidth]{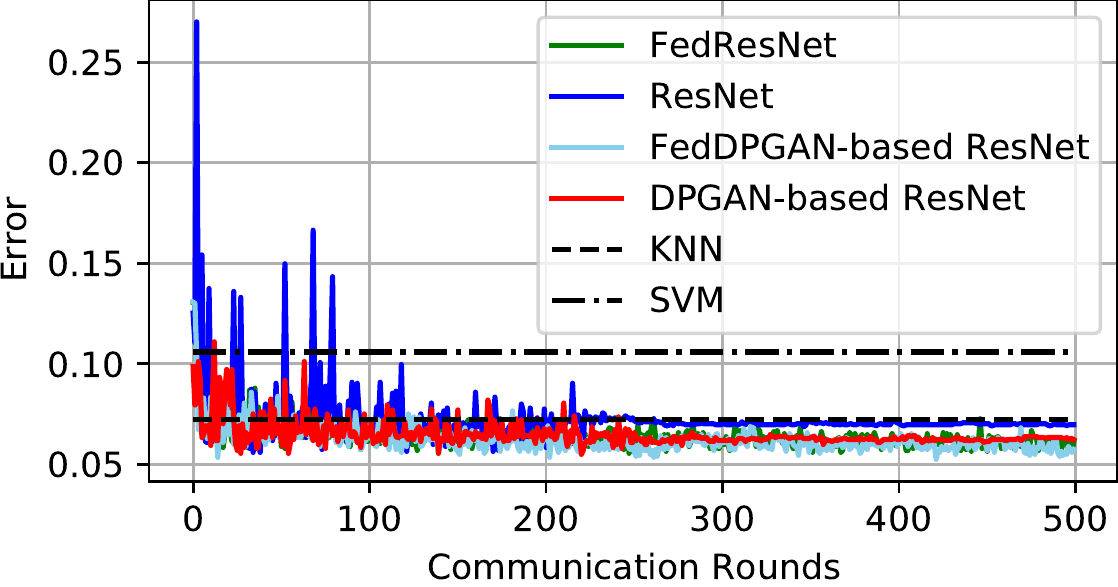}
			\label{fig3-b}}
		\caption{Comparison of COVID-19 diagnostic performance between the proposed model and the benchmark models.}
		\label{fig3}
	\end{figure*}
	
	\subsection{Model Performance}
\textcolor{black}{We compare the model performance of the proposed FedDPGAN-based ResNet model with that of FedResNet, DPGAN-based ResNet, ResNet, CNN, MLP, KNN, and SVM models with the same simulation configuration. Among these seven competing methods, DPGAN-based FedResNet and FedResNet are federated models and the rest are centralized models. ResNet \citep{he2016deep} has a good performance on image classification tasks and become a widely used baseline model.} The SVM, KNN, and MLP models are popular ML models for general classification applications \citep{1641014}.
	
\textcolor{black}{Table. \ref{table:1} indicates the accuracy of model and the compared model in diagnosing COVID-19.} From all of the results in the table, we can see that the proposed model can not only protect privacy but also use data augmentation method to improve performance. \textcolor{black}{Fig. \ref{fig3-a} shows that the performance of DPGAN-based ResNet model is better than the best baseline method centralized ResNet by 0.84\%, which is 4.36\% higher than the worst centralized baseline method SVM and is better than the worst deep learning baseline method CNN which is 3.05\% below.} The reason is that: (1) The large amount of data generated enables ResNet models to learn more samples. (2) Model training with DPGAN makes ResNet more generalizable.
	
In federated learning, our model can achieve the best model performance which is 0.49\% higher than the baseline method DPGAN-based ResNet model, as shown in Fig.\ref{fig3-b}. In a word, FedDPGAN-based ResNet model can achieve accurate without compromising privacy.

	\begin{table*}
		\centering
		\caption{Comparison of COVID-19 diagnostic performance between the proposed model and the benchmark models under IID setting.}
		\begin{tabular}{@{}cccc@{}}
			\toprule
			Model           & Accuracy & Data Augmentation & Privacy Protection \\ \midrule
			\textbf{FedDPGAN-based ResNet} & \textbf{ 94.45\%}    &\textbf{ \checkmark} & \textbf{\checkmark}\\
			FedResNet       & 93.96\%    & ×                 & \checkmark         \\
			DPGAN-based ResNet    & 93.77\%    & \checkmark        & \checkmark         \\
			ResNet \cite{he2016deep}          & 92.93\%    & ×                 & ×                  \\
			CNN \cite{tajbakhsh2016convolutional}            & 90.72\%    & ×                 & ×                  \\
			MLP  \cite{li2018novel}           & 92.05\%    & ×                 & ×                  \\
			KNN  \cite{park2018privacy}           & 92.78\%    & ×                 & ×                  \\
			SVM   \cite{morra2009comparison}          & 89.41\%    & ×                 & ×                  \\ \bottomrule
		\end{tabular}
		\label{table:1}
	\end{table*}

	\subsection{Performance of Federated Learning with Data Augmentation under IID and non-IID Settings}
\begin{figure*}
		\centering
		\large
		\subfigure []{\includegraphics[width=0.45\linewidth]{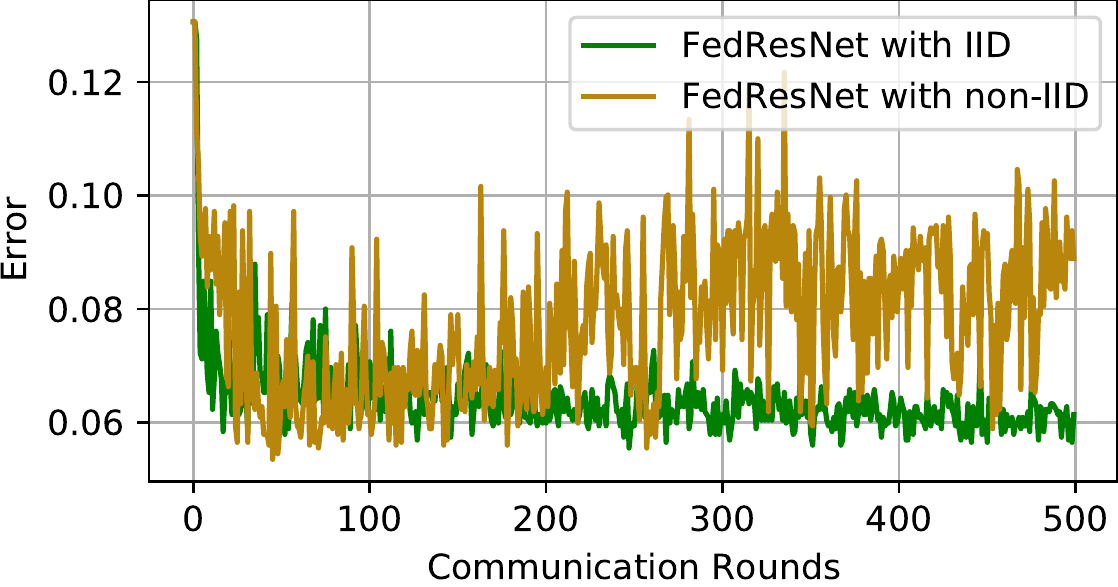}
			\label{fig4-a}}
		\hfill
		\subfigure[]{ \includegraphics[width=0.45\linewidth]{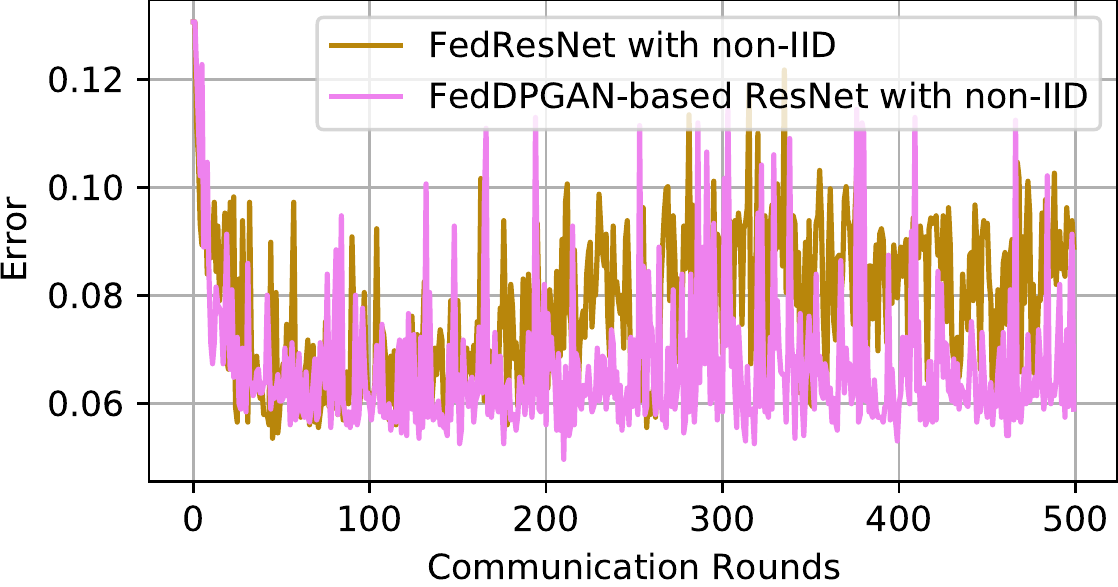}
			\label{fig4-b}}
		\caption{(a) Performance comparison of FedResNet model under IID and non-IID settings; (b) Performance comparison between FedResNet model and FedDPGAN-based ResNet model under non-IID settings.}
		\label{fig4}
	\end{figure*}

In this part, we quest the influence of data augmentation methods in IID and non-IID settings. First, we compare the performance of the FedResNet model under IID and non-IID settings. \textcolor{black}{Fig. \ref{fig4-a} shows that the prediction error of FedResNet model under the non-IID setting is 2.75\% higher than under the IID setting.} Experimental results show that non-IID distribution will affect the convergence performance of the model, resulting in the degradation of model. The reason is that the distribution of non-IID data will affect the convergence of the model, resulting in a decline in model performance.
	
\textcolor{black}{Second, under non-IID settings, we make overall evaluation of FedResNet by using data augmentation method and the FedResNet model without this method.} Fig.\ref{fig4-b} shows that the prediction error of FedDPGAN-based ResNet (using data augmentation method) model is 3.00\% lower than FedResNet (without using data augmentation method) under non-IID setting. The reason is that data augmentation methods can alleviate non-IID problems by generating diverse data. Such a method can make the convergence of federated learning training more stable.
	
\begin{figure*}
		\centering
		\large
		\subfigure []{\includegraphics[width=0.45\linewidth]{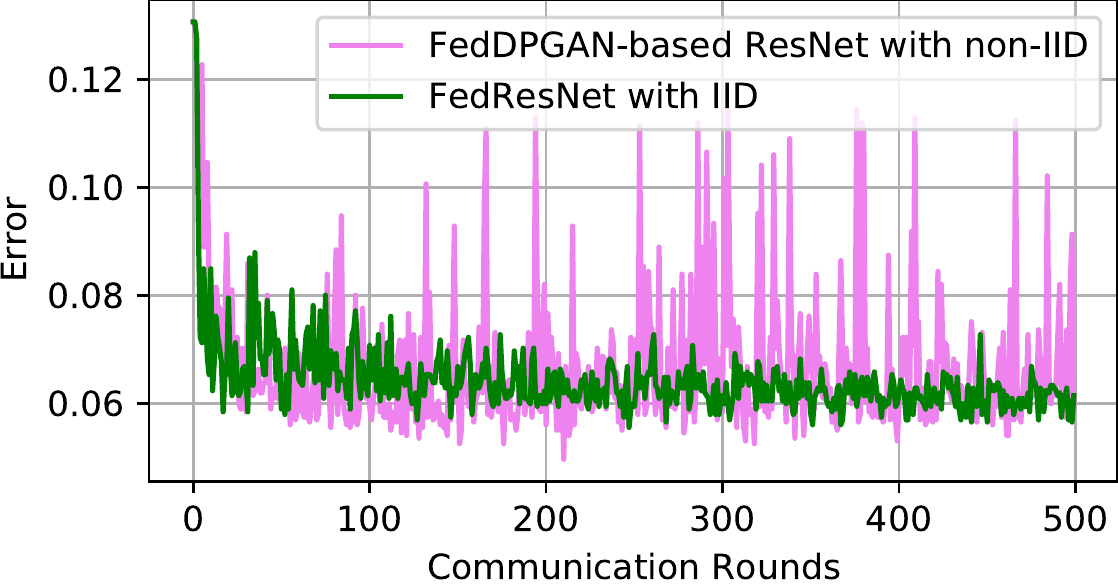}
			\label{fig5-a}}
		\hfill
		\subfigure[]{ \includegraphics[width=0.45\linewidth]{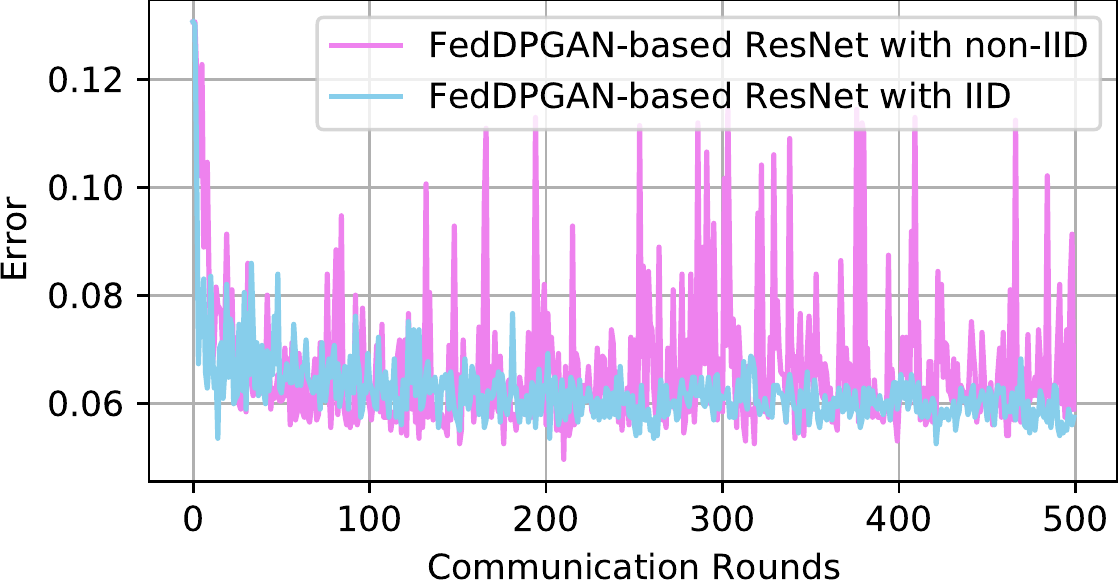}
			\label{fig5-b}}
		\caption{(a) Performance comparison of FedResNet model under IID setting and FedDPGAN-based ResNet model under non-IID settings; (b) Performance comparison between FedDPGAN-based ResNet model under IID setting and under non-IID settings.}
		\label{fig5}
	\end{figure*}

\begin{figure}
		\centering
		\large
		\includegraphics[width=0.95\linewidth]{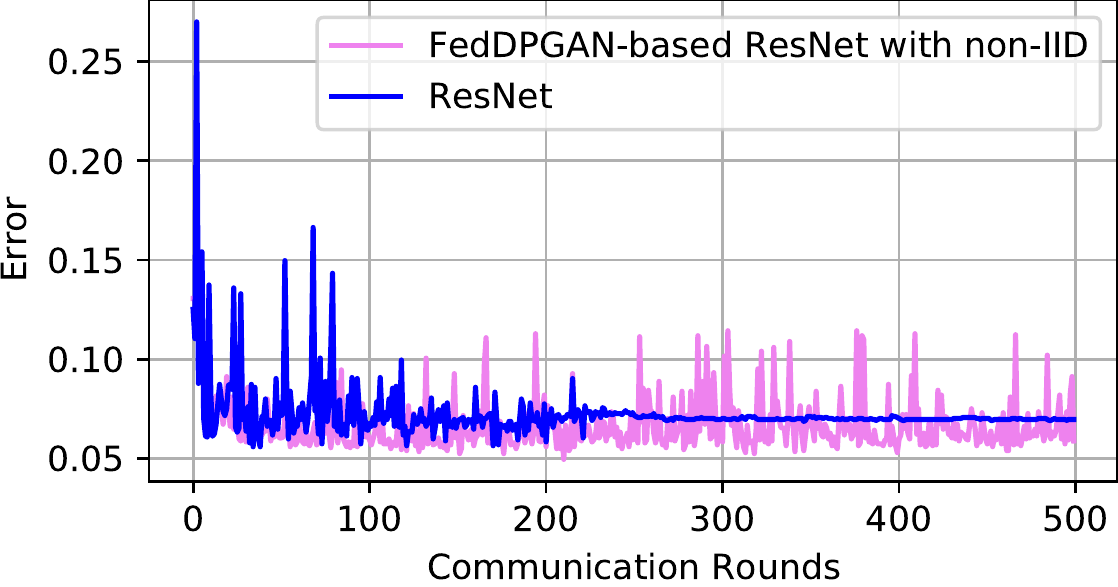}
		\hfill
		\caption{Performance comparison between ResNet model and FedDPGAN-based ResNet model under non-IID settings.}
		\label{fig6}
	\end{figure}

Third, we compare the performance of FedDPGAN-based ResNet model under non-IID setting and FedResNet model under IID setting. Fig. \ref{fig5-a} shows that the performance of the FedDPGAN-based ResNet model with non-IID setting is close to the FedResNet with IID setting. Furthermore, Fig. \ref{fig5-b} shows the prediction error of the proposed models under IID and non-IID settings. In this case, our model is superior to the centralized ResNet which without privacy protection as shown in Fig. \ref{fig6}. In a word, our model is more suitable for real-world medical application scenarios.
	
	\subsection{Performance Comparison of Different Privacy Budgets}
	
\textcolor{black}{In this part, we assess the performance of the proposed model by setting different privacy budgets.} Since the privacy budget $\sigma$ indicates the scale of Gaussian noise \citep{fredrikson2015model}, we need to explore the relationship between the scale of Gaussian noise and the performance of the proposed model. Specifically, we generate fake images of different quality by adjusting privacy budgets $\sigma$ and then explore their performance. We enhance the privacy protection ability of datasets through improving $\sigma$. From Table \ref{table:2}, we can draw a conclusion that the smaller size of $\sigma$ the higher model performance we will gain. \textcolor{black}{The experimental results show that we can adjust the privacy budget $\sigma$ to achieve a balance between performance and privacy protection.}
	
\begin{table}
		\centering
		\caption{Performance under the IID setting and non-IID setting of the proposed model under different privacy budgets $\sigma$.}
		\begin{tabular}{@{}cccccc@{}}
			\toprule
			Model           & Data distribution & $\sigma$     & Accuracy & Data Augmentation         & Privacy Protection           \\ \midrule
			FedDPGAN-based ResNet & non-IID              & $10^{-4}$ & 94.11\%  & \checkmark & \checkmark \\
			& non-IID              & $10^{-2}$ & 94.06\%  & \checkmark & \checkmark \\
			& non-IID              & $1$ & 91.90\%  & \checkmark & \checkmark \\
			\midrule
			FedDPGAN-based ResNet & IID                  & $10^{-4}$ & 94.45\%  & \checkmark & \checkmark \\
			& IID                  & $10^{-2}$ & 93.81\%  & \checkmark & \checkmark \\
			& IID                  & $1$ & 94.01\%  & \checkmark & \checkmark \\
			\bottomrule
		\end{tabular}
		\label{table:2}
\end{table}
	
\section{RESULTS AND DISCUSSION}
\textcolor{black}{We propose the FedDPGAN model can be used in diagnosing COVID-19 under using CXR images without compromising privacy.} Such a model enables hospitals in different geographic locations to collaboratively train a COVID-19 diagnostic model without sharing data. Specifically, our method solves two serious challenges currently encountered in diagnosing COVID-19: data availability and data privacy. First, in this model, we design a distributed DPGAN model to address data availability issue by generating COVID-19 image data. In particular, we use $(\varepsilon,\delta)$-DP noise to protect the privacy of GAN' training gradient. Second, we introduce FL framework to protect patient's privacy and apply the ResNet model to diagnostic COVID-19. \textcolor{black}{In the experiment part, we test the performance of FedDPGAN model on COVID-19 chest X-ray image datasets and compare it with centralized ResNet, CNN, MLP, KNN, and SVM models.} The results show that our method has the best model performance and privacy protection ability compared with competing methods. \textcolor{black}{Furthermore, the experimental results indicate that the GAN component in the proposed model can alleviate the non-IID problem in FL, which opens a window for the use of data augmentation to solve the non-IID problem.}
	
In the future, we will design a more realistic semi-supervised federated learning system \citep{liu2020rc} to solve the lack of data labeling and data privacy issues in the medical field. Furthermore, we will explore how data augmentation methods can improve the non-IID problem in FL, which motivates us to design more efficient data augmentation methods to solve non-IID problem in the future.


%
%

\bibliographystyle{spbasic}      
\bibliography{cas-refs}   

%
%

\end{document}